\documentclass[twocolumn,showpacs,preprintnumbers,amsmath,amssymb]{revtex4}

\usepackage{graphicx}
\usepackage{dcolumn}
\usepackage{bm}

\begin{document}

\title{Electromodulation of the Magnetoresistance in Diluted Magnetic Semiconductors Based
Heterostructures.}

\author{M.P.L\'opez-Sancho, M.C.Mu\~noz and L.Brey}
\affiliation{\centerline{Instituto de Ciencia de Materiales de Madrid (CSIC),~Cantoblanco,~28049 ~Madrid,~Spain.}}

\begin{abstract}
We study the properties of heterostructures formed by two layers 
of diluted magnetic semiconductor  separated by a nonmagnetic semiconductor layer. 
We find that 
there is a RKKY-type  exchange coupling  between the magnetic layers that oscillates between
ferromagnetic and antiferromagnetic  as
a function of the different parameters in the problem.
The different transport properties of these phases make that this heterostructure presents 
strong magnetoresistive effects.
The coupling can be also modified by an  electric field. We propose
that it is possible to alter dramatically the electrical resistance of the  heterostructure
by applying an electric field. 
Our results indicate that  in a single gated
sample
the magnetoresistance could be modulated by with an electrical   bias voltage.

\end{abstract}

\pacs{75.50.Pp, 75.70.Cn, 75.70.Pa}

\maketitle

Recently it has been possible to grow Mn doped GaAs semiconductors with a ferromagnetic
paramagnetic transition temperature, $T_c$,  near 100K\cite{Ohno}. Further experimental and theoretical
works indicate the possibility to obtain room temperature ferromagnetism in
others diluted Mn-doped III-V semiconductors\cite{Dietl-bis,konig1,Sonoda}.

Room temperature ferromagnetic semiconductors have a great potentiality to be  used
in magnetoelectronics and  spintronics. Therefore a big  effort is been doing in two directions:
first the study of the origen of ferromagnetism
and the search of high Curie temperature
magnetic semiconductors\cite{Dietl,Jungwirth1,Berciu,chatto,
Calderon1,Alvarez}, and second the design and growth of devices and heterostructures for
magnetoelectronics and spintronics\cite{Loureiro,Rossier,Brey,Jungwirth}.

The III-V high $T_c$ semiconductors have a concentration, $x$, of Mn 
ions randomly located. 
Experimentally, the optimal Mn concentration for obtaining high $T_c$'s is 
near $x \sim$0.06\cite{Potashnik}, and 
the magnetic impurities are rather diluted, hence the name Diluted Magnetic Semiconductors (DMS).
At small Mn concentration, each ion substitutes a column-III cation (III$_{1-x}$Mn$_x$-V), gets a
$S$=5/2 local moment and gives a hole to the host semiconductor.
A large amount of these holes are trapped on antisite n-type deep defects present in the
host semiconductor, since  these materials are grown at low temperatures\cite{Potashnik}.
The rest of holes, with a concentration $p<<x$, are responsible for the occurrence
of ferromagnetic order in DMS. 
The system is formed by two interacting subsystems: a subsystem of Mn ions which are so
dilute that direct interaction between their magnetic moment is negligible, and
a subsystem of carriers. 
Without interaction between them, both systems should  be paramagnetic at any temperature. 
However, the antiferromagnetic Hund's coupling, $J$, between the carrier and the Mn spins 
makes the two subsystems
to become ferromagnetically ordered and antiferromagnetically coupled. 
The Mn ions feel a long range ferromagnetic interaction 
mediated
by the itinerant spin polarized carriers\cite{Dietl-bis,konig1,Sonoda}.
The Curie temperature of the DMS's  depends on the carrier density of states at
the Fermi energy, the Hund's coupling and the density of magnetic impurities.

In this work we study the properties of heterostructures formed by two slabs  of thickness
$d_M$  of Mn-based DMS separated by a nonmagnetic semiconductor layer of thickness $d_P$. Analyzing 
this problem in the framework of the mean field approximation, we investigate  the electrical, magnetic and
transport properties of the heterostructures  as  function of $x$, $p$, $d_M$, $d_P$, Hund's
coupling, band offset and external bias voltage.

The main conclusions of this work are the following:
\par \noindent
i) There is a RKKY-type  exchange interaction between the magnetic layers that oscillates in sign as
a function of the different parameters in the problem.
Positive and negative signs correspond to  antiferromagnetic (AF) and  ferromagnetic (F)
coupling between the magnetic slabs.
\par \noindent
ii)Assuming that the alloy scattering is much smaller in the paramagnetic semiconductor than
in the Mn doped layers, we find that the electrical resistivity, for currents flowing
parallel to the interfaces, has rather different values for the F and 
AF 
coupled heterostructures.
\par \noindent 
iii) We predict that 
the electrical resistance can be modulated
not only  with an external magnetic field, but also by applying 
and external electric field.

DMS's are described by the following Hamiltonian, 
\begin{equation}
H\!=\!H_{h}\!+\! J\!  \sum _{I,i}\! {\bf S} _ I  \cdot {\bf s} _{i} \delta
({\bf r} _i\! -\! {\bf R}_I) \!+\!  
\,
W \! \!\sum  _{I,i}  n_I  \delta
({\bf r} _i\! -\! {\bf R}_I)   . 
\label{Hamiltonian1}
\end{equation}    
Here  $H_{h}$ 
describes the carriers, it is the sum of the kinetic energy of the holes and the hole-hole
interaction energy. For the range of carrier density of interest in DMS's, 
the carrier-carrier interaction is not relevant and  we neglect it. 
We treat the kinetic energy in the framework of the envelope function approximation. In this
approach we describe the hole electronic states of the host semiconductor 
by a parabolic
band. For GaAs, the effective mass, $m^*$ is 0.5$m_e$\cite{konig1}.
The carriers are confined in the whole heterostructure and their motion  is restricted to 
$-d_M-d_P/2 < z < d_M + d _P / 2$.
The last two terms  in Eq.(\ref{Hamiltonian1})  represent the coupling between the  carriers and the 
Mn's. The term proportional to $J$  is the antiferromagnetic 
exchange interaction between the spin $ {\bf S} _ I$ of the Mn$^{2+}$ ions located
at ${\bf R}_I$ and the spins, ${\bf s} _{i}$ of the itinerant carriers. This term is  responsible
for  the long range ferromagnetic interaction between the Mn core spins. 
The last term in Eq.(\ref{Hamiltonian1})
is an interaction between the carrier charge, $n _i$ and the potential arising from the magnetic 
dopants. The origin of $W$ is the different electronegativity of Mn and GaAs atoms, and the
screening of the Mn by the carriers. The value of $J$ is typically 
0.1-0.15$ eVnm^3$\cite{Matsukura,Matsukura-bis,Omiya,Okabayashi}. Since there is not reliable
experimental information on the value of $W$, we consider it as a parameter with value $0<W<J$.
The direct magnetic interaction between the Mn ion spins is considerably 
weaker than the interaction
with the carrier spins and therefore we neglect it. 

We solve Hamiltonian (\ref{Hamiltonian1}) in the mean field approximation. In this 
approach, similar to the Jellium model, the local magnetic interaction of the spin
carriers with the Mn spins is substituted by the interaction with an effective magnetic field 
of intensity\cite{nota2} $SJx/a^3$ and  directed parallel to the Mn's ion spin polarization.
In the same spirit of the Jellium model, the electronegativity difference  between the 
carriers and the Mn ions 
is described by an effective potential of interaction $Wx /a ^3 $.
In these expressions 
$a^3$ is the unit cell volume of  the host semiconductor.

\begin{figure}
\includegraphics [clip,width=8.cm,height=9.cm]{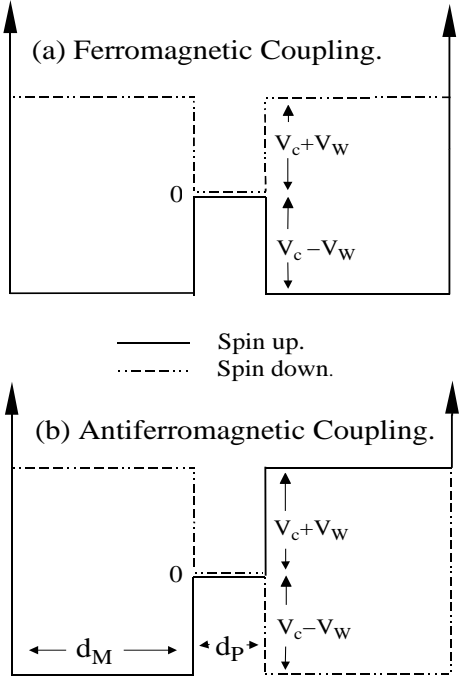}
\caption{
Potential profiles for both spin carriers directions, in the F and AF 
configurations.}
\label{esquema}
\end{figure}               

Within this approach, in our heterostructure the carriers are free to move in the $x-y$ plane
and the one particle wave functions  and eigenvalues have the form
\begin{equation}
\Psi ^{\alpha} _ {i, {\bf k} _{\perp}, \sigma} = {\frac{e ^{i {\bf k} _{\perp} {\bf r} _{\perp}}
}{\sqrt{S}}} 
\Phi ^{\alpha} _ {i,  \sigma} (z) \, \, \, , \, \, \, 
\varepsilon  ^{\alpha} _ {i, {\bf k} _{\perp}, \sigma} =  { \frac{\hbar k ^2 _{\perp}} 
{2 m ^* }} + \varepsilon  ^{\alpha} _ {i, \sigma}
\, \, \, \, \, . 
\label{eigen}
\end{equation}
Here $S$ is the areal dimension of the sample, ${\bf r} _{\perp}$ and
${\bf k} _{\perp}$ are the  position and the momentum of the carriers 
in the plane perpendicular to the growth direction, $i$ is a subband index and  $\Phi ^{\alpha} _ {i,  \sigma} (z)$ and
$\varepsilon  ^{\alpha} _ {i, \sigma}$ are obtained from the  one-dimensional
Schrodinger equation,
\begin{equation}
\left (  - {\frac{ \hbar ^2 }{2  m ^ *}} { \frac{d ^2}{d z ^2}} + V _{\sigma} ^{\alpha} (z)
\right )  \Phi ^{\alpha} _ {i,  \sigma} (z) = \varepsilon  ^{\alpha} _ {i, \sigma}
\Phi ^{\alpha} _ {i,  \sigma} (z) \,  \, \, \, ,
\label{Hamiltonian_mf}
\end{equation}
where $\sigma$ is the carriers spin index, up (+) or down (-), $\alpha$  stands for the solutions
with  
ferromagnetic (F)   or antiferromagnetic  coupling (AF) between the DMS layers, and the effective potential $ V _{\sigma} ^{\alpha} (z)$
has the following form, see fig.(\ref{esquema})
\begin{equation}
V _{\pm} ^{\alpha} (z) =  \left\{
\begin{array}{cc}
\mp V_c+V_W & \mbox{if}     \, \, \,  -d_M-d_P/2 < z < -d_P/2 \\
0 & \mbox{if} \, \, \, \, \, \, \, \, \, \, \, \, \, \, \, \, \,  -d_P/2 < z < +d_P/2 \\
\mp C_{\alpha} V_c+V_W  & \mbox{if} \, \, \, \, \, \, \, \, \, \, \, \, \,  d_P/2 < z < d_M+ d_P/2 \\
\infty  & \mbox{otherwise}  
\end{array}
\right. 
\label{effective_pot}
\end{equation}
where $C_{\alpha}$ is +1(-1) in the F(AF) coupling case,  $V_c=SJx/(2a^3)$ and
$V_W=Wx/a^3$. 
By summing the energy of the occupied states, 
we obtain the total energy per unit area of the solutions with ferromagnetic,$E_F$,  and
antiferromagnetic, $E_{AF}$, coupling between the GaMnAs layers.

\begin{figure}
\includegraphics [clip,width=8.cm]{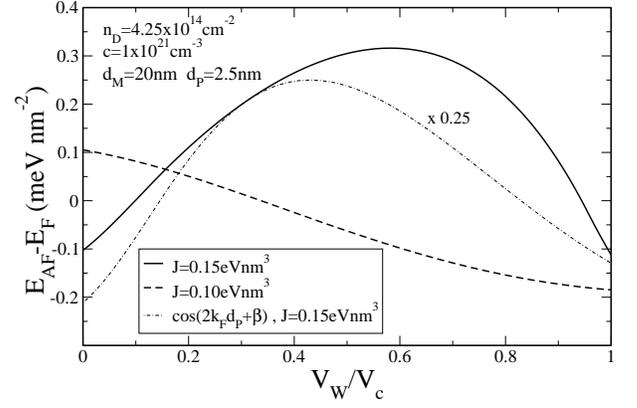}
\caption{Energy difference between the AF and the F coupled magnetic layers cases as a function
of the bandoffset. 
We also plot $\cos(2k_Fd_P+\beta)$ to show  the RKKY origin of the coupling.}
\label{figura1}
\end{figure}               

In fig.(\ref{figura1}) we plot the difference $E_{AF}$-$E_F$ as a function of the 
band-offset, $V_W$,  for two  values of $J$. The parameters of the heterostructure are
$d_M$=20$nm$ and  $d_P$=2.5 $nm$,  we consider that the DMS layers have  a density of Mn's, 
$c$=1$\times$10$^{21}cm^{-3}$, and the two-dimensional density of carriers is
$n_{D}$=4.25$\times$10$^{14}cm^{-2}$, that  roughly corresponds to a three dimensional density
of carriers ten times smaller then  the density of magnetic impurities.
The exchange coupling oscillates as a function of the band-offset. The coupling is due to a RKKY-like 
interaction between the DMS layers similar to that occurring in magnetic multilayers\cite{libro}.
The coupling exists because there is  charge carriers  in the whole heterostructure.
In particular in the central region, there is a paramagnetic hole gas which mediates
the interaction between the magnetic slabs.
To enlighten that, we have plotted the quantity $\cos(2k_Fd_P+\beta)$, being $k_F$ the 
Fermi wave-vector  of the  paramagnetic hole gas in the
central layer, and $\beta$ an arbitrary phase.  From  the comparison between the numerical
results and the cosine, we conclude that the  RKKY model accounts for the existence 
of the exchange coupling oscillations.
The value of the phase $\beta$ has been chosen to make  the comparison easier.
We have studied the  coupling as a function of $W$, but we
have also found
oscillations in the exchange energy by changing
other parameters in the heterostructure which alters the product $2k_Fd_P$.
Similar oscillations in DMS based superlattices have been reported in ref.(\cite{Jungwirth}).
Also,  indications of interlayer exchange coupling have been observed
in GaMnAs/GaAs superlattices\cite{Mathieu}.
For DMS based heterostructures the value of the exchange coupling energy is smaller than in metallic systems. However 
the magnetic field necessary to overcome the AF coupling is
$B \sim \frac {E_F-E_{AF}}{g \mu _B S c d _M}\sim$ 100$-$1000Gauss, large
enough for magnetoresistive applications.

The possibility of changing the layer coupling from AF to F, by applying a magnetic field
imply that the heterostructure should present large  magnetoresistance (MR) i.e.   
change of the electrical resistance when applying a magnetic field.
The transport parallel to the growth direction should present a MR near 100$\%$.
However the use of this geometry would imply tunneling processes to inject  the carriers 
inside and outside
the heterostructure.

\begin{figure}
\includegraphics [clip,width=8.cm]{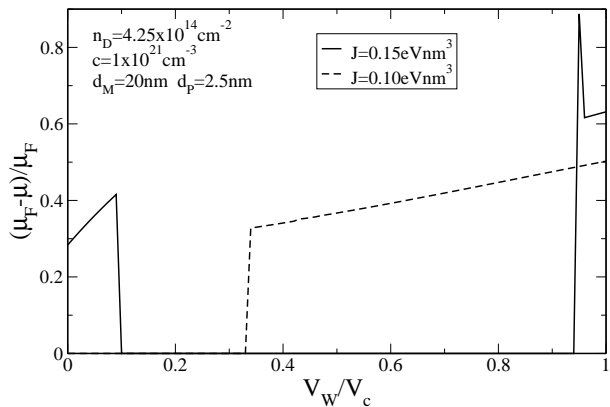}
\caption{Absolute value of MR  as a function of $V_W$. 
}
\label{figura2}
\end{figure}               

Transport parallel to the interfaces also presents large  MR. We
calculate the conductivity, $\mu$,  of the heterostructure assuming that the main source of
scattering are the impurities and the antisite defects  located in the DMS 
layers. Assuming point contact interaction between the carriers
and the impurities\cite{nota3}, the following expression for $\mu$  is 
obtained\cite{bastardbook}
\begin{equation}
\mu = \frac {e ^2}{m^*} \sum _{i,\sigma} \tau _i ^{\sigma} \frac {n ^ {\sigma} _i}{n_{D}}
\, \, \, \, \, ,
\label{conductivity}
\end{equation}
where $n ^ {\sigma} _i$ is the two dimensional density of carriers in the 
subband $i$ and spin $\sigma$,  the sum is over the occupied states and
the transport scattering time is given by,
\begin{equation}
\frac{1}{\tau_i ^{\sigma}} = C \frac{1}{\sum _ l {A_{i,\sigma,l,\sigma}}}\, \, \, 
\label{scattering_time}
\end{equation}
with the sum   restricted to the occupied states, $C$ is a constant which
depends on the details of the scattering potential  and
\begin{equation}
A_{i,\sigma,l,\sigma}= \int _{imp. \, region} dz 
\left | \Phi   _ {i, \sigma} (z) \right | ^2
\left | \Phi   _ {l, \sigma} (z)\right | ^2
\, \, \, \, \, .
\label{solapes}
\end{equation}
This integral is restricted to the DMS regions. 

We have evaluated the conductivity of the F and AF solutions and in fig.(\ref{figura2})
it is plotted the absolute value of the MR as a function of the band offset
for the cases shown in fig.(\ref{figura1}). MR  is different from cero, with
values bigger than 20$\%$, 
when the ground state is antiferromagnetic.
In the F case the minority carriers are mainly localized in the central paramagnetic
layer, where the scattering is much weaker than in the DMS layers. In the AF
coupling case the carriers, for both  spin orientation, located on the central layer have a 
wavefunction extended on one of the DMS slabs  and therefore they suffer a stronger
scattering. The minority spin high mobility channels in the F phase, 
localized in the central layer, are responsible 
for the conductivity difference  between the F and the AF phases. 
The peak that appears near $V_W \sim$0.95$V_C$ is a quantum effect due to the occupancy
of a new subband  (see Eq.(\ref{scattering_time}))
in the F phase.

\begin{figure}
\includegraphics [clip,width=8.cm]{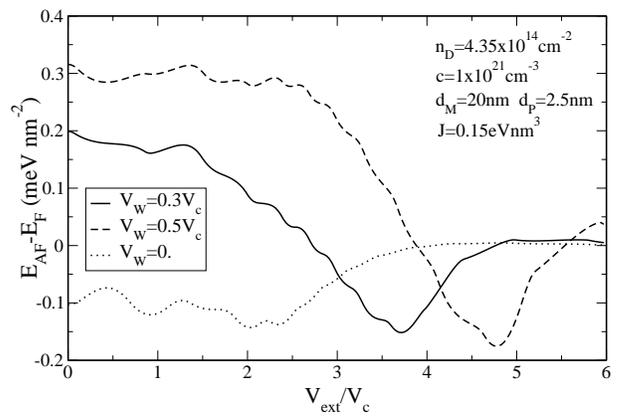}
\caption{Exchange energy coupling  as a function of $V_{ext}$.
}
\label{figura3}
\end{figure}               

Because semiconductors have  lower carrier density than metals,  the
exchange coupling is sensitive to moderate external electrical bias, $V_{ext}$ applied
from left to right  of the heterostructure.
In fig.(\ref{figura3}) we plot the exchange energy coupling as a function of $V_{ext}$, for
different values of $V_W$. The application of the bias changes 
the charge density in the central layers and therefore the coupling oscillates and changes sign.
In the cases of $V_W$=0.3 and 0.6$V_C$ the bias changes the coupling from F to AF.
At large enough bias one of the DMS layers becomes almost  depopulated and the 
two magnetic slabs   become decoupled.
There is also an superimposed  fine structure  related to the depopulation
of the subbands originated by $V_{ext}$. 

\begin{figure}
\includegraphics [clip,width=8.cm]{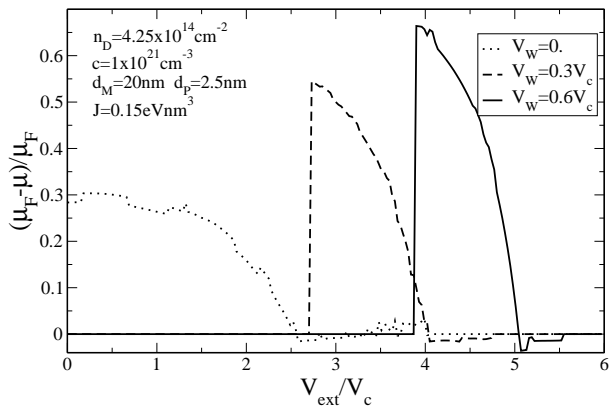}
\caption{Absolute value of the MR as a function of $V_{ext}$.}
\label{figura4}
\end{figure}

In fig.(\ref{figura4}) we plot the absolute value of the MR as a function of $V_{ext}$. Because the
bias changes the coupling from F to AF, or viceversa, there is 
a big change in the MR as a function of $V_{ext}$. As  for the energy coupling there
are small oscillations, related with the subbands occupation,  superimposed on the mean structure.
The main point of the results presented in fig.(\ref{figura4}) is the large 
change in the resistance of the heterostructure when a bias is applied.
For the parameter values corresponding to DMS, the magnitude of the
electric field necessary to electromodulate the MR is $\sim$100$KV/cm$
that is an experimentally reasonable value.

The results shown in figures (\ref{figura3}) and (\ref{figura4}) are not
only interesting for possible applications in magnetoresistive devices, but
also for the basic study of quantum phase transitions.  It could be 
possible to see a quantum phase transition between the  F and the  AF phases
in a single gated sample by varying the bias voltage.

In conclusion we have studied heterostructures formed by two layers of DMS separated by a 
nonmagnetic semiconductor. We find that there is a RKKY-like interaction between the layers
that oscillates in sign with the parameters of the problem, and with an 
applied electrical field. The ferromagnetic and antiferromagnetic coupled heterostructures
have rather different resistances and  present  large  magnetoresistive
effects. We predict that, 
for a fixed heterostructure, an external bias changes the coupling
from ferromagnetic to antiferromagnetic, producing a big variation on the electrical resistance
and magnetoresistance. Therefore the resistance can be modulated, not just with an external
magnetic field, but also
by applying and external bias voltage.

Financial
support is acknowledged from Grants No PB96-0085, BFM2000-1330, BFM2000-1107 (MEC, Spain) and Fundaci\'on Ram\'on Areces.



\begin{thebibliography}{25}
\bibitem[{not({\natexlab{a}})}]{nota2}
\bibinfo{note}{In the interaction between the carrier spins and the efective
  magnetic field, created by the Mn's we take $g \mu _B =1 $}.
\bibitem[{not({\natexlab{b}})}]{nota3}
\bibinfo{note}{We only consider the scattering of the carriers by the
  electrostatic part of the Mn's impurity potential. At low temperatures
  spin-flip scattering is considerably weaker. See for example
  ref(\cite{Calderon_sw})}.
\end{thebibliography}

\end{document}